\documentclass[sigchi]{acmart}

\usepackage{booktabs} 

\usepackage{adjustbox}

\begin{document}
\settopmatter{printacmref=false} 
\renewcommand\footnotetextcopyrightpermission[1]{} 
\pagestyle{plain} 
\title{Chatbots as Conversational Recommender Systems in Urban Contexts}

\author{Pavel Kucherbaev}
\affiliation{%
  \institution{Delft University of Technology}
  \streetaddress{Mekelweg 4}
  \city{Delft} 
  \state{the Netherlands} 
  \postcode{2628 CD}
}
\email{p.kucherbaev@tudelft.nl}

\author{Achilleas Psyllidis}
\affiliation{%
  \institution{Delft University of Technology}
  \streetaddress{Mekelweg 4}
  \city{Delft} 
  \state{the Netherlands} 
  \postcode{2628 CD}
}
\email{a.psyllidis@tudelft.nl}

\author{Alessandro Bozzon}
\affiliation{%
  \institution{Delft University of Technology}
  \streetaddress{Mekelweg 4}
  \city{Delft} 
  \state{the Netherlands} 
  \postcode{2628 CD}
}
\email{a.bozzon@tudelft.nl}

\renewcommand{\shortauthors}{P. Kucherbaev et al.}

\begin{abstract}
In this paper, we outline the vision of chatbots that facilitate the interaction between citizens and policy-makers at the city scale. We report the results of a co-design session attended by more than 60 participants. We give an outlook of how some challenges associated with such chatbot systems could be addressed in the future.
\end{abstract}



\maketitle

\section{Introduction}
Messenger applications, such as Messenger, Telegram, Whatsapp and WeChat, represent a popular medium of communication, which people use to interact with friends, family members and various brands. In 2015 the total number of active users of such applications surpassed the total number of users of conventional social network applications \cite{BIIntelligence}. Chatbots, on the other hand, are computer programs living in messenger applications and emulating a conversation with a human to provide a certain service \cite{McTearCallejasGriol16}. Conversational recommender systems derive user preferences by conversing with users in natural language \cite{Christakopoulou2016}. There are several examples of using chatbots in urban contexts, such as for pedestrian navigation \cite{Janarthanam2013} 
and policy decisions \cite{Augello2009,Boden2006}. While these examples serve users in very specific use cases, our vision is to develop a chatbot system that provides various recommendations to people in city regions and, further, helps citizens interact with the municipality.

While the open data movement is spreading wide, the majority of municipal datasets remain siloed, making it difficult for people to use. Solutions provided by municipalities are usually implemented in the form of standalone applications or web portals, which citizens are usually reluctant to install or visit. With the proposed chatbot system we help citizens (a) receive recommendations, based on open data sources (e.g. good kindergartens, parking lot locations etc.); (b) provide information to municipal authorities (e.g. to report a pothole); and (c) directly communicate with municipality employees or with other citizens, if the requested information is not available in any data source. To understand if such a chatbot system is needed, to derive requirements and collect use cases that both citizens and policy-makers want to see in it, we organised a co-design workshop.

\section{Co-design Workshop}
We organised a workshop on February 17, 2017 at \emph{Pakhuis de Zwijger} in Amsterdam, which is a platform for discussing city-related topics. More than 60 people with diverse backgrounds (e.g. students, industry employees, researchers, municipality employees), representing various nationalities, attended the session. The age of the participants ranged from 20 to 75 years, and around 40\% were women. About 50\% of the audience expressed their prior familiarity with chatbots. We wanted to brainstorm with the participants about possible use cases and requirements of chatbot systems for urban data retrieval and recommendation. To give them an idea of what is possible to do with chatbots, we created a mockup\footnote{More info on the co-design workshop is available at: \url{http://bit.ly/2uwuujO}} showing how citizens can ask information and recommendations about Amsterdam, provide extra information to the chatbot system, and even be asked proactively by it. Table 1 summarises the examples developed together with our participants. 
Later, we asked people in the audience to pick any use case and come up with a paper mockup of the conversation with the system.

Participants worked on their mockups in 10 teams of 5-7 people. The co-design session allowed us to identify a number of domains where current solutions do not yet satisfy users well. At the same time, surprisingly, the navigation use case, which is discussed a lot in the literature \cite{Janarthanam2013, Kashioka2011}, did not arise during the session, suggesting that this particular use case is already well covered by other systems (e.g. GoogleMaps) or that participants did not consider it appropriate for a conversational interface. There is a big room for solutions recommending places (e.g for spending free time), and presenting municipal, governmental, and neighborhood-related information (e.g. planned interventions, trash bin status etc.).

\begin{table}
\caption{Chatbot usage ideas derived from the co-design workshop.}
\begin{adjustbox}{width=\columnwidth}
\footnotesize
\begin{tabular}{ll}
\multicolumn{1}{c}{\textbf{Citizens can ask the bot about}}                 & \multicolumn{1}{c}{\textbf{Citizens can provide the bot with}}         \\ \hline
\multicolumn{1}{|l|}{What is being built here?}                             & \multicolumn{1}{l|}{Report a pothole}                   \\ \hline
\multicolumn{1}{|l|}{What is the construction schedule?}                    & \multicolumn{1}{l|}{Report children-friendly places}     \\ \hline
\multicolumn{1}{|l|}{Where is the best place to buy X?}                     & \multicolumn{1}{l|}{Report full trash cans}            \\ \hline
\multicolumn{1}{|l|}{What is the closest open night market?}                & \multicolumn{1}{l|}{Report things available for free}   \\ \hline
\multicolumn{1}{|l|}{What is the best time to visit the office?} & \multicolumn{1}{l|}{Report sentiment and feelings}    \\ \hline
\multicolumn{1}{|l|}{What is a good place to dance?}                        & \multicolumn{1}{l|}{Report about neighbourhoods}  \\ \hline

\end{tabular}
\end{adjustbox}
\end{table}

\section{Outlook}
\begin{figure}[h]
\includegraphics[width=1.0\columnwidth]{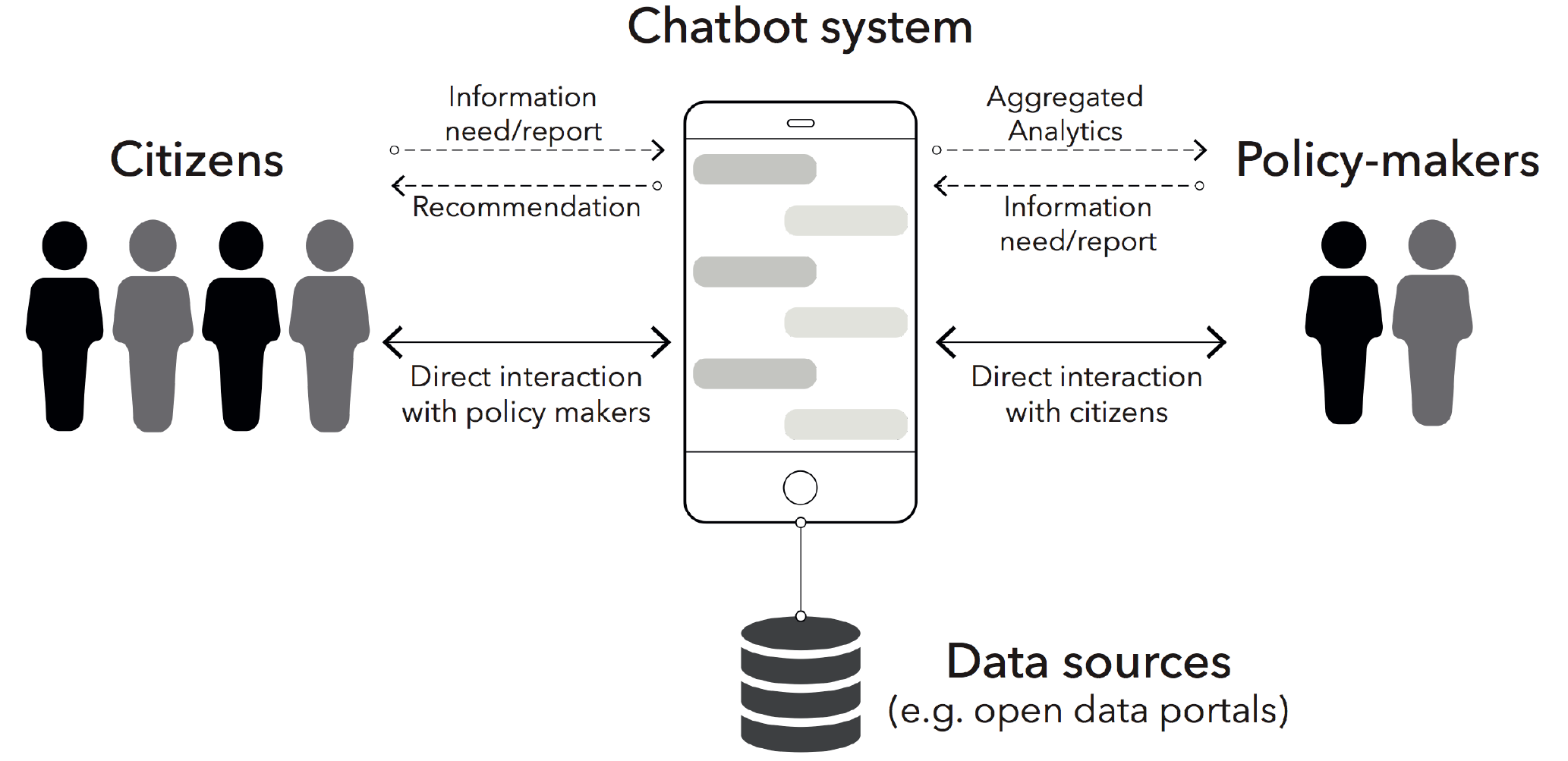}
\centering
\caption{Conceptual schema of the chatbot system.}
\label{schema}
\end{figure}

Taking into consideration the use cases from the co-design session, we introduce a conceptual schema of the interactions citizens and policy-makers are engaged in through the chatbot system (Figure \ref{schema}). We envision that \emph{citizens} can request information via the chatbot system from connected \emph{data sources} (e.g. municipal open data portals). Moreover, citizens can proactively report new information, which will be stored in associated data sources. \emph{Policy makers} can as well request relevant information for decision-making purposes. If it is not possible to automatically satisfy an information need of a citizen, a responsible municipality employee is assigned by the chatbot system, such that the citizen can interact with this employee through the chatbot directly. Similarly, some citizens can express their willingness to be contacted by policy makers to provide their feedback to new policies and regulations. 

Developing such a chatbot system is not trivial, but there are various approaches that make it possible to obtain comparable results today. \emph{Human computation} could be used to address limitations of fully automatic chatbots, such when they do not understand user requests and, therefore, cannot provide a useful response. 
\emph{Gamification} strategies can be implemented to motivate citizens to collect and report information about the city.

We are confident that such a chatbot system can significantly improve the civic engagement of citizens, making it easy to contribute to the city and to be involved in discussions about urban issues. The accessibility of such a chatbot system is very high, as existing messaging systems facilitate the on-boarding of citizens and minimize the learning curve. 
While implementing such a chatbot system is much easier in cities which are in the forefront of urban development, the system could be transferred to cities of developing countries to facilitate the interaction of people with city stakeholders.



\begin{acks}
We thank the participants of the workshop and the personnel of Pakhuis de Zwijger. The research is supported by the Amsterdam Institute for Advanced Metropolitan Solutions  with the \emph{AMS Social Bot} grant.
\end{acks}

\bibliographystyle{ACM-Reference-Format}
\bibliography{sigproc} 

\end{document}